\documentclass[onecollarge,natbib]{svjour2}
\bibpunct{[}{]}{;}{n}{}{,} 
\smartqed  
\usepackage{graphicx}
\usepackage{color}
\usepackage{amsmath}
\journalname{Few-Body Systems (EFB23)}

\newcommand{\bm}[1]{ \mbox{\boldmath $#1$}  }

\begin{document}

\title{Strongly Interacting One-Dimensional Systems with Small Mass Imbalance
}


\author{Artem~G.~Volosniev 
}


\institute{Artem~G.~Volosniev  \at
                  Institut f{\"u}r Kernphysik, Technische Universit{\"a}t Darmstadt, 64289 Darmstadt, Germany \\
 \email{volosniev@theorie.ikp.physik.tu-darmstadt.de}
}


\date{Received: date / Accepted: date}

\maketitle

\begin{abstract}
We study a strongly interacting system of $N$ identical bosons and one impurity in a one-dimensional trap.  First, we assume that the particles have identical masses and analyze the corresponding set-up. After that, we study the influence of a small mass asymmetry on our analysis. In particular, we discuss how the structure of the wave function and the degeneracy in the impenetrable regime depend on the mass ratio and the shape of the trapping potential. To illustrate our findings, we consider a four-body system in a box and in an oscillator. We show that in the former case the system has the smallest energy when a heavy (light) impurity is close to the edge (center) of the trap. And we demonstrate that the opposite is true in the latter case.
\keywords{one spatial dimension \and mass imbalanced systems \and Tonks-Girardeau gas}
\end{abstract}

\section{Introduction}
\label{intro}

It is known that multicomponent impenetrable one-dimensional systems fermionize and can be analyzed analytically if the masses of particles are the same \cite{girardeau, girardeau2007, deuret2008}. This solution can be used to understand some features of the eigenstates even if the interaction is finitely strong \cite{volosniev2014, deuret2014, cui2014, harshman2014, levinsen2015, volosniev2015, gharashi2015, guan2015}. The obtained information has been used to explain experimental results \cite{serwane, zurn, murmann2016}, as well as to analyze the related numerical calculations \cite{gharashi, Astrakharchik2013, lindgren, grining2015, bellotti2016}. However, the situation changes when the masses of constituents differ. In this case exact solutions exist only for a few specific mass ratios, see, e.g, Refs.~\cite{mcguire1963,olschanii2015}, and, therefore, one needs to develop other approaches in order to analyze the system. Fortunately, for few-body systems one can simply adopt known numerical methods, see, e.g., studies in Refs.~\cite{pecak2015, amin2016, pecak2016, molte2016, miguel2016}.  One important observation of these studies is that there is a separation of the components in the strongly interacting regime driven by the mass ratio (cf.~Ref.~\cite{Cui2013}). This result is very interesting, in particular, because it suggests an additional tool to engineer magnetic correlations with cold atoms. In this paper we study these dynamics for a small mass imbalance using perturbation theory. For simplicity we consider a system of bosons with a single impurity, however, the discussion can be easily generalized to more than one impurity.

The paper is organized as follows. In Sec.~\ref{sec:1} we present the Hamiltonian. In Sec.~\ref{sec:2} we address the equal mass case. Here we write the spin chain Hamiltonian that describes a strongly interacting system and discuss it in the thermodynamic limit. In Sec.~\ref{sec:3} we consider a system with a small mass imbalance and use perturbation theory to estimate its energy. Finally, in Sec.~\ref{sec:4} we conclude with a short summary of the results obtained in the paper.   

\section{Formulation of the problem}
\label{sec:1}

We consider a one-dimensional system that consists of $N$ identical spinless bosons, each of mass $m$, and a distinguishable particle of mass $M$, which we name the impurity. All the particles are placed along the $x$-axis, the former are at the coordinates $x_1,...,x_N$ and the latter is at $x_{N+1}\equiv s$. We assume that the bosons interact with each other via the potential $V_{bb}(x_i-x_j)=\kappa g\delta(x_i-x_j)$ and with the impurity via $V_{bi}(x_i-s)=g \delta(x_i-s)$, where $\delta(x)$ is the Dirac delta function, and the parameters $g,\kappa > 0$ characterize the interaction strengths. Additionally, we assume that the system is confined by some external potential, $V_{ext}$, which is the same for all particles.

This system is described by the Hamiltonian
\begin{equation}
\label{hamilt}
H=-\frac{\hbar^2}{2m}\sum_{i=1}^N\frac{\partial^2}{\partial x_i^2}+\sum_{i=1}^NV_{ext}(x_i)+\kappa g\sum_{i>j}\delta(x_i-x_j)-\frac{\hbar^2}{2M}\frac{\partial^2}{\partial s^2}+V_{ext}(s)+g\sum_{i=1}^N\delta(x_i-s).
\end{equation}
In the present paper we examine it assuming that the interaction energy $g\delta(x)$ sets the largest energy scale of the problem. 
Note that this set-up with $m=M$ is well-studied, see, e.g., Ref.~\cite{zinner2016} and references therein. We in turn are interested in what happens to the system when $M=(1+\xi)m$ and $|\xi|\ll 1$. Our starting point is the known solution in the equal mass case that we then 'correct' using the methods of perturbation theory. It is wortwhile noting that systems with $|\xi|\ll 1$ can be realized with alkali atoms in the isotopic mixtures, e.g., in $^{40}$K -$^{41}$K systems (cf. Ref.~\cite{wu2011}), or in $^6$Li - $^7$Li (cf. Ref.~\cite{barbut2014}).

\section{Equal mass case}
\label{sec:2}
Let us first discuss the system with $m=M$, assuming that one particle can still be singled out for the role of the impurity, e.g., it can have another spin projection.  We start with $1/g=0$. In this limit the particles cannot exchange their positions, thus, the  configurations\footnote{Here the configuration $IB...B$ denotes the part of coordinate space where all the bosons sit to the left of the impurity, i.e., $s<x_{1,2,...,N}$, $BI...B$ means that $N-1$ bosons are to the left of the impurity and one is to the right, etc.} $IB...B$, $BI...B$, ..., $BB...I$ are not coupled to one another, i.e., if we place the impurity, for instance, to the left of the bosons then it will stay there forever. Note that these configurations have identical energies as once we work with a specific ordering then the statistics is not important: we should simply solve the Schr{\"o}dinger equation for $N+1$ impenetrable particles of the same mass. This observation means that, for example, the ground state is always $(N+1)$-fold degenerate. 

If $1/g\neq 0$ then the particles can exchange their positions, hence, a specific ordering starts to couple to the rest. This coupling (in leading order) is decribed by the following spin chain Hamiltonian \cite{volosniev2014, deuret2014, volosniev2015}
\begin{equation}
H_s=E \mathbf{I} -\sum_{j=1}^{N}
\left [\frac{J_j}{2g}\left({\bm \sigma}^j {\bm \sigma}^{j+1}-\mathbf{I}\right)-
\frac{J_j}{g\kappa}\left(\sigma_z^j \sigma_z^{j+1}+\mathbf{I}\right)\right ],
\label{ham_spin}
\end{equation}
where $E$ is the energy of the system at $1/g=0$, $\mathbf{I}$ is the identity matrix, ${\bm \sigma}^j=(\sigma^j_x,\sigma^j_y,\sigma^j_z)$ are the Pauli matrices at site $j$, and the coupling coefficient $J_j$ is
\begin{equation}
J_j =  -\frac{(N+1)!}{g}\int_{a<x_1<...<x_N<b} 
\prod_{k=1}^{N}\mathrm{d}x_k \left[\dfrac{\partial \Phi(x_1,...,x_N,s)}{\partial s}
\right]_{s=x_j}^2,
\end{equation} 
here $\Phi$ is the normalized wave function of $N+1$ trapped spinless fermions, i.e., $\Phi$ is the Slater determinant made up of the one-body wave functions $\phi_i(x)$ in the potential $V_{ext}(x)$. The interval $[a,b]$ is the domain of $\phi_i(x)$. Note also that for simplicity here and below we use the system of units in which $\hbar=m=1$.

To understand the meaning of $J_i$ we calculate the average value for the system in the ground state, i.e., 
\begin{equation}
J\equiv\frac{1}{N}\sum_{i=1}^N J_i =-\frac{N+1}{g}\int_a^b \mathrm{d}x_1 ... \int_a^b \mathrm{d}x_N \left[\frac{\partial \Phi_{ground}}{\partial s}\right]^2_{s=x_1}.
\end{equation}
This expression can be written as the sum of one-dimensional integrals
\begin{equation}
J=-\frac{1}{g N} \sum_{\underset{i\neq k}{k,i=1}}^{N+1} \int\mathrm{d}x_1 \left\{\frac{\mathrm{d}\phi^*_i(x_1)}{\mathrm{d}x_1} \frac{\mathrm{d}\phi_i(x_1)}{\mathrm{d}x_1}\phi^*_k(x_1)\phi_k(x_1) + \frac{\mathrm{d}\phi^*_i(x_1)}{\mathrm{d}x_1} \phi_i \frac{\mathrm{d}\phi_k(x_1)}{\mathrm{d}x_1} \phi^*_k \right\}.
\label{eq:Javerage}
\end{equation}
Let us evaluate these integrals for the homogeneous system, i.e.,  $V_{ext}=0$, in the thermodynamic limit, i.e., $N\to\infty$, $(b-a)\to\infty$ and $N/(b-a) \to \rho$. To that end,  we use $\phi_i(x)=\frac{e^{ik_i x}}{\sqrt{b-a}}$, $k_i\in\{\pm 2\pi n/(b-a)\}$, where $n$ is an integer number. We neglect pieces that vanish in the thermodynamic limit and obtain
\begin{equation}
J\simeq -\frac{1}{g}\frac{8\pi^2}{(b-a)^3}  \sum_{i=1}^{N/2} i^2 = -\frac{\pi^2}{3 g}\left(\frac{N}{b-a}\right)^3,
\label{eq:Javeragethermodynamic}
\end{equation}
here for simplicity we have assumed that $N$ is even. We see that $J=-2 \epsilon/\gamma$, where $\gamma=g/\rho \gg 1$ is the dimensionless interaction strength and $\epsilon=\rho^2\pi^2/6$ is the energy per particle in the ground state, i.e., $\epsilon\equiv E_{ground}/N$. Therefore, for fixed $\gamma$, $J$ is determined by $\epsilon$, cf. Ref.~\cite{matveev2004}. Note that this information together with the density from the local density approximation\footnote{The local density approximation assumes that $n(x)=\sqrt{2(\mu-V_{ext}(x))}/\pi$, where $\mu$ is the chemical potential.} can be used to estimate the values of $J_i$ also for finite non-homogeneous systems \cite{deuret2014, marchukov2016, li2016}. 

The discussion above allows us to understand also the dependence of $J$ on temperature around absolute zero, i.e., at $T\ll k_B/\epsilon$ ($k_B$ is the Boltzmann constant) and $\rho = \mathrm{const}$. Indeed, in this case the temperature-dependent coupling coefficient is $J_T=-2 \epsilon_T/\gamma$, where $\epsilon_T$ is the energy per particle at $T$, so
\begin{equation}
J_T\simeq J\left(1+\left(\frac{k_B T}{\rho^2 \pi^2}\right)^2\frac{\pi^2+1}{2}\right).
\end{equation}
This expression implies that in the trap $V_{ext}(x)$ the temperature effects are inhomogeneous. For instance, in a weakly varying\footnote{A weak variation here means a vanishingly small change of the trap on the length scale given by $1/n$.} trap it is smaller in denser regions, i.e., where the energy per particle at $T=0$ is larger. 

Finally, we use Eq.~(\ref{eq:Javerage}) to calculate $J_j$ in a box, as it was noted \cite{marchukov2016} that in such a trap $J_j=J$. We obtain
\begin{equation}
J=-\frac{1}{g N} \sum_{i\neq k} \int\mathrm{d}x \left\{k_i^2 \cos^2(k_i x) \sin^2(k_k x)  + k_i k_k \cos(k_i x) \sin(k_i x) \cos(k_k x) \sin(k_k x)  \right\}.
\end{equation}
This integral can be easily evaluated. For example, for the ground state we have
\begin{equation}
J=-\frac{\pi^2 }{g(b-a)^3}\sum_{i=1}^{N+1} i^2 = -\frac{\pi^2 (N+1)(N+2)(2N+3)}{6 g (b-a)^3}.
\end{equation}
Note that in the thermodynamic limit this equation reproduces the expression in Eq.~(\ref{eq:Javeragethermodynamic}).

\section{Mass imbalanced case}
\label{sec:3}

Let us start with the impenetrable case, i.e., $1/g=0$. In this limit the particles 
preserve their relative positions, hence, as before, we can study the system by examining different configurations, i.e., $IB...B$, $BI...B$, ..., $BB...I$. The first thing to notice here is that these orderings do not necessarily have the same energy if $m\neq M$, and, thus, not all orderings may be included in the adiabatic ground state.  Let us illustrate this observation with a simple example: a very heavy impurity in a box.  In this case, if we put the heavy particle at one of the box edges we effectively eliminate it from the problem, and, hence, end up with the smallest possible energy. Therefore, we expect that the ground state in our example contains only the $IB...B$ and $BB...I$ structures. This situation should be compared with the mass-balanced case, in which all the configurations are populated \cite{volosniev2014}. The comparison suggests that by changing the mass ratio the structure of the wave function changes. Let us investigate this change in more detail. 
To this end, we assume that $m/M\to 1$, i.e., $|(M-m)/m|=|\xi|\ll 1$, and calculate perturbative corrections to the eigen states at $\xi=0$. The perturbation is due to the operator 
\begin{equation}
V_{pert}\equiv -\frac{1}{2(1+\xi)}\frac{\partial^2}{\partial s^2} +\frac{1}{2}\frac{\partial^2}{\partial s^2}= \frac{\xi}{2(1+\xi)}\frac{\partial^2}{\partial s^2},
\end{equation}
and, hence, the first order correction to the energy is
\begin{equation}
\Delta E_i = (N+1)! \frac{\xi}{2} \int_{a<x_1<x_2<...<x_{i-1}<s<x_i<...<x_N<b}\Phi\frac{\partial^2 \Phi}{\partial s^2}.
\label{eq:energy_corr}
\end{equation}
Here the subscript $i$ indicates the ordering to which we calculate the energy correction, i.e., $\Delta E_1$ is the energy correction to the ordering $IB...B$, $\Delta E_2$ is to $BI...B$ etc. In general, in a trap all $\Delta E_i$ will be different, therefore, the spectrum of a mass-imbalanced system in the impenetrable limit is not degenerate\footnote{Of course, we cannot rule out some accidental degeneracies or degeneracies due to intrinsic symmetries of the trapping potential, e.g., under the parity transformation, see the examples below.} unless the mass ratio is one.  Note also that the integral in Eq.~(\ref{eq:energy_corr}) is the kinetic energy of the impurity in a particular configuration, therefore, the sign of the correction is fully determined by $\xi$, i.e., a heavy (light) impurity always decreases (increases) the energy. Accordingly, the configuration that has the largest (smallest) value of $|\Delta E_i|$ is the ground state of a system with a heavy (light) impurity.

To understand the meaning of these corrections we estimate their average in the thermodynamic limit for $V_{ext}=0$, i.e., we calculate
\begin{equation}
\Delta E \equiv \frac{1}{N+1}\sum_{i=1}^{N+1} \Delta E_i = -\xi \frac{E}{N+1}.
\end{equation}
In the thermodynamic limit this expression leads to $\Delta E = -\xi \epsilon$, and implies that in an inhomogeneus system with a weakly changing density the energy is minimal if a heavy (light) impurity sits in the densest (sparsest) region. This behavior is intuitively expected, as in the densest region the kinetic energy of a particle is the largest, therefore by placing a heavy particle there we maximally reduce the energy. At the same time, in the sparcest region the kinetic energy of a particle is the smallest, therefore by placing a light impurity there we minimally increase the energy. 
Naturally, the thermodynamic limit cannot be used as a reference point for few-body systems. Therefore, let us illustrate our findings by calculating $\Delta E_i$ for four-body systems in a box and in a harmonic oscillator. 

\vspace*{2em}

{\bf System in a box.} The one-body wave functions in a box have to vanish at $x=a$ and $x=b$, and, thus, they must have the form
\begin{equation}
\phi_i(x)=\sqrt{\frac{2}{b}}\sin\left(\frac{i \pi x}{b}\right),
\end{equation} 
here for simplicity we have used $a=0$.
We utilize these functions with $i=1,2,3,4$ to construct the Slater determinant in Eq.~(\ref{eq:energy_corr}) that in turn yields the corrections to the ground state energies 
\begin{equation}
\Delta E^{gr}_{1,4}= -4.05\xi \frac{\pi^2}{b^2}, \qquad \Delta E^{gr}_{2,3} = -3.44 \xi \frac{\pi^2}{b^2}.
\end{equation}
First thing to notice here is that the degeneracy present in the equal mass case is partially lifted now. For example, the ground state is doubly and not 4-fold degenerate. These corrections also imply that a heavy impurity ($\xi>0$) minimizes the energy when it is located close the edge of the trap\footnote{We expect that the state $(IBBB+BBBI)/\sqrt{2}$ is adiabatically connected to the lowest state at $1/g \neq 0$, because it is symmetric under the parity transformation.} (cf.~Ref.~\cite{pecak2015}), just as an infinitely heavy impurity from the example above. A system with a light impurity ($\xi<0$) exhibits the opposite behavior. Note that these dynamics cannot be obtained using the results that we derived in the thermodynamic limit. Indeed, if we use the density from the local density approximation we will obtain that all $\Delta E_i$ are identical. This result is not surprising, as the local density approximation is not supposed to work well at the edges of the box, i.e., in the region where we expect the heavy impurity to be.   Finally, let us consider the excited manifold in which $i=1,2,3,5$ single-particle orbitals are populated. In this case $\Delta E^{exc}_{1,4}=-4.85 \xi \pi^2 /b^2$ and $\Delta E^{exc}_{2,3} = - 4.89 \xi \pi^2 /b^2$, and, therefore, the situation here is reversed, e.g., a heavy impurity minimizes the energy when sits in the middle. 

\vspace*{2em}

{\bf System in a harmonic oscillator.} Consider now a harmonic oscillator with the characterstic length $l$, i.e., $V_{ext}(x)=x^2/2l^4$. The corresponding one body wave functions are 
\begin{equation}
\phi_i(x)=\frac{1}{\sqrt{2^i \; i! \; l}}e^{-\frac{x^2}{2 l^2}}H_i\left(\frac{x}{l}\right),
\end{equation}
where the functions $H_i(x)$ are the physicists' Hermite polynomials. These functions with $i=1,2,3,4$ allow us to construct $\Phi$ in Eq.~(\ref{eq:energy_corr}), and, hence, to obtain the corrections to the ground state energies:
\begin{equation}
\Delta E_{1,4} = -0.75 \frac{\xi}{l^2}, \qquad \Delta E_{2,3} = -1.24 \frac{\xi}{l^2}.
\end{equation}
These expressions suggest that the ground state of a system with a heavy impurity is doubly degenerate. It consists of the configurations $BIBB$ and $BBIB$, the state $(BIBB+BBIB)/\sqrt{2}$ is correspondingly the adiabatic ground state. This result is consistent with what we expect from the thermodynamic limit, i.e., a heavy impurity lowers the energy when it is located in the densest region. It also agrees with the results derived for harmonically trapped systems in Refs.~\cite{pecak2016, loft2015}. Note, however, that in these studies the external potential depends on mass.

\vspace*{2em}

Let us now briefly discuss the system with $1/g\neq 0$. Again we can describe it using the Heisenberg Hamiltonian in which the information from Eq.~(\ref{eq:energy_corr}) is included in a magnetic term, i.e., 
\begin{equation}
H_s=E \mathbf{I} -\sum_{j=1}^{N}
\left [\frac{J_j}{2g}\left({\bm \sigma}^j {\bm \sigma}^{j+1}-\mathbf{I}\right)-
\frac{J_j}{g\kappa}\left(\sigma_z^j \sigma_z^{j+1}+\mathbf{I}\right)\right]+\frac{1}{2}\sum_{i=1}^{N+1} \Delta E_i \left(\sigma_z^i+\mathbf{I}\right).
\label{ham_spin_mass_imb}
\end{equation}
We see that if $J/g \sim \Delta E$ then all the terms here are important and should be taken into account. In this regime the parameter $\xi/g$ gives us an additional tool to engineer inhomogeneous Heisenberg Hamiltonians suitable, for example, for state transfer applications cf. \cite{bose2003, marchukov2016tr}. It also can give us additional insight into dynamics of a heavy (light) impurity in the Tonks-Girardeau gas (cf. \cite{gamayun2014,visuri2016}).

\section{Summary}
\label{sec:4}

In this paper we investigated a bosonic system with an impurity. We assumed that the interactions are strong and studied what happens if the mass of the impurity differs slightly from that of a boson. However, first, in Sec.~\ref{sec:2} we discussed the equal mass case, see also \cite{volosniev2014, deuret2014}. Here we derived Eq.~(\ref{eq:Javerage}) for the average value of the coefficients that allowed us to study the system in the thermodynamic limit. It is worthwhile noting that this
equation can be useful in numerical calculations of $J_i$ \cite{deuret2016, loft2016}, e.g., to control the accuracy or to compute one of the elements. Also it can be used to find the lowest energy directly, e.g., if $\kappa=1$. Indeed, in this case the ground state wave function is simply $|\Phi_{ground}|$ and the corresponding energy is $E + 2 J N$. 

We presented the results for the mass-imbalanced case in Sec.~\ref{sec:3}. Here using perturbation theory we showed that in a box potential a heavy impurity minimizes the ground state energy when placed at the edge of the trap, whereas in a harmonic trap the opposite is true, cf.~Ref.~\cite{pecak2015}.  Furthermore, we considered the system in the thermodynamic limit, and argued that in the systems for which the local density approximation is valid a heavy impurity minimizes the energy by staying close to the global minimum of the potential. 

\vspace*{1em}

{\small {\bf Acknowledgments} The author is indebted to Nikolaj Zinner for his inspiring comments and for many helpful discussions. The author thanks Lasse Kristensen, Niels Jakob Loft, Oleksandr Marchukov, and Manuel Valiente for their remarks on the manuscript. The author gratefully acknowledges the support of the Humboldt Foundation.}


\end{document}